\documentclass[12pt]{article}
\usepackage[dvips]{epsfig}

\newcommand{\beq}{\begin{equation}}
\newcommand{\eeq}{\end{equation}}

\begin{document}

\title{Deformed solitons: The case of two coupled scalar fields}
\author{A. de Souza Dutra \and UNESP/Campus de Guaratinguet\'{a}-DFQ}
\maketitle

\begin{abstract}
In this work, we present a general procedure, which is able to generate new
exact solitonic models in 1+1 dimensions, from a known one, consisting of
two coupled scalar fields. An interesting consequence of the method, is that
of the appearing of nontrivial extensions, where the deformed systems
presents other BPS solitons than that appearing in the original model.
Finally we take a particular example, in order to check the above mentioned
features.
\end{abstract}

\newpage

Despite of being less usual than linear systems, the nonlinear ones and
particularly those having solitonic excitations are very interesting and
important in order to modelling many physical, biological and chemical
systems. A very important example is that of the electrical conductivity of
some organic materials, where polarons and other polymer chain solitons were
responsible for the appearing of conducting polymers \cite{nobel}. Another
important appearance of solitons is that related to electrical conduction
through DNA molecules \cite{dna}. In the literature there exists a great
number of models and applications of the solitonic solutions of one and two
dimensional kind \cite{nature}, \cite{dionisio}.

In a recent paper, Bazeia et al \cite{deformed} introduced a method capable
of generating new exact solitonic systems from known ones, for the case of a
scalar field in 1+1 dimensions. In this work, we intend to study the
difficulties and restrictions of a naive generalization of the cited
approach, and then present a general manner of getting new exact solitonic
models from a known one, this time for the case of two or more coupled
scalar fields.

The Lagrangian density for the the case of two coupled scalar fields which
we are going to work with is given by
\begin{equation}
\mathcal{L}=\frac{1}{2}\left( \partial _{\mu }\phi \right) ^{2}+\frac{1}{2}%
\left( \partial _{\mu }\chi \right) ^{2}-\,V\left( \phi ,\chi \right) ,
\end{equation}

\noindent whose Euler-Lagrange equations in 1+1 dimensions are given by
\begin{equation}
\frac{d^{2}\phi \left( x\right) }{dx^{2}}=W_{\phi }W_{\phi \phi }+W_{\chi
}W_{\chi \phi }\,,\,\frac{d^{2}\chi \left( x\right) }{dx^{2}}=W_{\chi
}W_{\chi \chi }+W_{\phi }W_{\phi \chi }\,,\,
\end{equation}

\noindent where we particularized the potential to a class which can be
written in terms of a superpotential $W$, as
\begin{equation}
\,V\left( \phi ,\chi \right) =\frac{1}{2}\,W_{\phi }^{2}+\frac{1}{2}%
\,W_{\chi }^{2}\,,
\end{equation}

\noindent and $W_{\phi }$ and $W_{\chi }$ stands for, respectively, the
differentiation with respect to the fields appearing in the lower index. For
this class of systems, one can show that the minimum energy solutions can be
obtained from the equivalent system of coupled first-order differential
equations \cite{bags}
\begin{equation}
\frac{d\phi }{dx}=\,W_{\phi }\left( \phi ,\chi \right) ,\,\frac{d\chi }{dx}%
=\,W_{\chi }\left( \gamma ,\chi \right) .
\end{equation}

If one starts from the above differential equations, and recover the
corresponding second-order ones, one gets
\begin{equation}
\frac{d^{2}\phi \left( x\right) }{dx^{2}}=W_{\phi }W_{\phi \phi }+W_{\chi
}W_{\phi \chi }\,,\,\frac{d^{2}\chi \left( x\right) }{dx^{2}}=W_{\chi
}W_{\chi \chi }+W_{\phi }W_{\chi \phi }\,,
\end{equation}

\noindent which are identical to those coming from the Euler-Lagrange equation as
written above, provided that the superpotential be twice differentiable. In other
words, the following restriction shows up

\begin{equation}
W_{\phi \chi }=W_{\chi \phi }.
\end{equation}

The energy of the so called $BPS$ states can be calculated straightforwardly, giving
\begin{equation}
E_{B}=\frac{1}{2}\int_{-\infty }^{\infty }dx\left[ \left( \frac{d\phi }{dx}%
-W_{\phi }\right) ^{2}+\left( \frac{d\chi }{dx}-W_{\chi }\right)
^{2}+W_{\chi }\frac{d\chi }{dx}+W_{\phi }\frac{d\phi }{dx}\right] ,
\end{equation}

\noindent from which we can see that the minimal energy will come from the
solutions obeying the following set of first-order differential equations
\begin{equation}
\frac{d\phi }{dx}=W_{\phi }\,;\,\frac{d\chi }{dx}+W_{\chi }
\end{equation}
and the energy of the field configuration is finally given by

\begin{equation}
E_{B}=|W\left( \phi _{i},\chi _{i}\right) -W\left( \phi _{j},\chi
_{j}\right) |,
\end{equation}

\noindent where $\phi _{i}$ and $\chi _{i}$ are the $i$th vacuum state of
the model \cite{junction}.

At this point we perform a general transformation in the fields
\begin{equation}
\phi =f\left( \theta ,\varphi \right) ,\,\chi =\,g\left( \theta ,\varphi
\right) ,
\end{equation}

\noindent which after some simple manipulations lead us to
\begin{equation}
\frac{d\theta }{dx}=\,W_{\theta }\left( \theta ,\varphi \right) \,,\,\frac{%
d\chi }{dx}=\,W_{\varphi }\left( \theta ,\varphi \right) ,
\end{equation}

\noindent where we defined
\begin{equation}
\,W_{\theta }\left( \theta ,\varphi \right) =\frac{1}{J}\left( \frac{%
\partial g}{\partial \theta }\,W_{\phi }\,\left( \theta ,\varphi \right) -%
\frac{\partial f}{\partial \theta }W_{\chi }\,\left( \theta ,\varphi \right)
\right) ,
\end{equation}

\noindent and
\begin{equation}
W_{\varphi }\left( \theta ,\varphi \right) =\frac{1}{J}\left( \frac{\partial
f}{\partial \varphi }\,W_{\chi }\,\left( \theta ,\varphi \right) -\frac{%
\partial g}{\partial \varphi }W_{\chi }\,\left( \theta ,\varphi \right)
\right) ,
\end{equation}

\noindent with the Jacobian of the transformation given as usually by
\begin{equation}
J\,\left( \theta ,\varphi \right) =\frac{\partial g}{\partial \theta }\,%
\frac{\partial f}{\partial \varphi }\,-\,\frac{\partial f}{\partial \theta }%
\,\frac{\partial g}{\partial \varphi }\,.
\end{equation}

Unfortunately however, the derivative of the superpotential appearing at the
right-hand side of the first-order equations does not obeys the crossed
derivative rule, which in its turns is strictly necessary to guarantee that
the solutions of the first-order equations are also solutions of the
corresponding second-order ones, as must happens when studying the so called
BPS solitons \cite{bps}. In order to become clearer the situation, we
exemplify the idea by studying a particular example inspired in one proposed
in the paper by Bazeia et al \cite{deformed}. Namely we have
\begin{equation}
\phi =\sinh \left( \varphi \right) \,,\,\chi =\theta .
\end{equation}

\noindent Furthermore we apply it to a well known model presenting solitonic solutions
\cite{bags},
\begin{equation}
\frac{d\phi }{dx}=\lambda \,\left( \phi ^{2}-a^{2}\right) +\mu \,\chi
^{2}\,,\,\frac{d\chi }{dx}=-2\,\mu \,\phi \,\chi ,  \label{S1}
\end{equation}

\noindent which, after performing the necessary calculations introduced
above, leaves us with the following set of equations
\begin{equation}
\frac{d\varphi }{dx}=\sec h\left( \varphi \right) \left[ \lambda \left( \sin
h\left( \varphi \right) ^{2}-a^{2}\right) +\mu \,\theta ^{2}\right] \,,\,%
\frac{d\theta }{dx}=-\,2\,\mu \,\theta \,\,\sin h\left( \varphi \right) \,.
\end{equation}

\noindent It is easy to verify that, indeed, by using the solutions of the
system
\begin{equation}
\phi =-a\,\tanh \left( 2\,\mu \,a\,x\right) ;\,\chi =\pm \,a\sqrt{\frac{%
\lambda }{\mu }-2}\,\sec h\left( 2\,\mu \,a\,x\right) ,
\end{equation}

\noindent one obtains the correct solution of the system of equations,
\begin{equation}
\varphi \left( x\right) =arc\sinh \left( -\,a\,\tan h\left( 2\,\mu
\,a\,x\right) \right) ;\,\theta \left( x\right) =\pm \,a\,\sqrt{\frac{%
\lambda }{\mu }-2}\,\sec h\left( 2\,\mu \,a\,x\right) .
\end{equation}

Notwithstanding, the above solutions of the first-order differential equations are not
solutions for the corresponding second-order ones. This happens precisely due to the
fact that $W_{\theta \varphi }\neq W_{\varphi \theta }$. From now on, we are going to
present an approach which is able to recover two new deformed nonlinear systems, from
the above ones, which accomplish with the conditions to be BPS states. For reach this
goal, we start by noting that the superpotential can be determined from each one of
the above equations, giving
\begin{equation}
W^{\left( 1\right) }\left( \theta ,\varphi \right) =\int \mathcal{D}\theta
\,\,W_{\theta }\left( \theta ,\varphi \right) \,+\,H^{(1)}\left( \varphi
\right) ,
\end{equation}

\noindent or
\begin{equation}
W^{\left( 2\right) }\left( \theta ,\varphi \right) =\int \mathcal{D}\varphi
\,W_{\varphi }\left( \theta ,\varphi \right) \,+\,H^{(2)}\left( \theta
\right) ,
\end{equation}

\noindent where $H^{(1)}\left( \varphi \right) $ and $H^{\left( 2\right)
}\left( \theta \right) $ are arbitrary functions which will be fixed in
order to guarantee that the condition $W_{\theta \varphi }^{\left( i\right)
}=W_{\varphi \theta }^{\left( i\right) }\,,\,\left( i=1,2\right) $ be
satisfied. Now, imposing that one of the solutions described in above
satisfies this condition, we obtain respectively
\begin{equation}
W_{\theta \varphi }\left( \theta ,\varphi \right) +H_{\varphi }^{(1)}\left(
\varphi \right) =W_{\varphi \theta }\left( \theta ,\varphi \right) \,,
\end{equation}

\noindent and
\begin{equation}
W_{\varphi \theta }\left( \theta ,\varphi \right) \,+\,H_{\theta
}^{(2)}\left( \theta \right) =W_{\theta \varphi }\left( \theta ,\varphi
\right) .
\end{equation}

The last step is to determine the arbitrary function $H_{\varphi }^{(1)}\left( \varphi
\right) $ or $\,H_{\theta }^{(2)}\left( \theta \right) $, by using our knowledged of
the relation between the original and the transformed fields, obtainable from the
inversion of the transformations,
\begin{equation}
\theta =f\left( \phi ,\chi \right) ,\,\varphi =\,g\left( \phi ,\chi \right) ,
\end{equation}
\noindent and also of the solutions of those original fields $\phi \left(
x\right) $ and $\chi \left( x\right) $. Then it is possible to write one
field in terms of the another one ($\theta =\theta \left( \varphi \right) $
or $\varphi =\varphi \left( \theta \right) $). So, one can finally discover
the expression of $H^{(1)}\left( \varphi \right) $ or $H^{(2)}\left( \theta
\right) $. In doing so, one can recover two systems of BPS equations:
\begin{equation}
\frac{d\theta }{dx}=\,W_{\theta }\left( \theta ,\varphi \right) \,;\,\frac{%
d\varphi }{dx}=\,W_{\varphi }\left( \theta ,\varphi \right) +H_{\varphi
}^{(1)}\left( \varphi \right) ,
\end{equation}

\noindent or
\begin{equation}
\frac{d\chi }{dx}=\,W_{\varphi }\left( \theta ,\varphi \right) \,;\,\frac{%
d\theta }{dx}=\,W_{\theta }\left( \theta ,\varphi \right) +H_{\theta
}^{(2)}\left( \theta \right) .
\end{equation}

Let us now present a concrete realization of the idea above presented. We
start by treating the case discussed here when we was showing that a naive
generalization of the idea outlined in \cite{deformed} does not works for
the construction of deformed solitons when two or more coupled fields are
present. Then, it is easy to verify that
\begin{equation}
\theta ^{2}\left( x\right) =\left( \frac{\lambda }{\mu }-2\right) \left(
a^{2}-\sinh \left( \varphi \right) ^{2}\right) ,
\end{equation}

\noindent and now imposing the requirement defined earlier in the text, one
obtains after straightforward calculations that
\begin{equation}
H_{\varphi }^{(1)}\left( \varphi \right) =\allowbreak \left( \sinh \varphi
^{2}-a^{2}\right) \left[ 2\mu \sec h\varphi +\left( \lambda -2\mu \right)
\cosh \varphi \right] ,
\end{equation}

\noindent whose corresponding set of coupled BPS equations are
\begin{equation}
\frac{d\theta }{dx}=-\,2\,\mu \,\theta \,\,senh\left( \varphi \right) ;\,
\end{equation}

\noindent and
\begin{equation}
\frac{d\varphi }{dx}=\mu \,\cosh \left( \varphi \right) \,\theta
^{2}\,+\left( \sin h\left( \varphi \right) ^{2}-a^{2}\right) \left[ 2\,\mu
\,\sec h\left( \varphi \right) +\left( \lambda -2\,\mu \right) \cosh \left(
\varphi \right) \right] .
\end{equation}

It is easy to verify now that this last system has the correct behavior as a
BPS one \cite{bps}. In other words, they generate a potential of the type
appearing in (4), coming from a superpotential given by

\begin{equation}
W_{1}=\mu \,\sinh \left( \varphi \right) \,\theta ^{2}+H_{1}\left( \varphi
\right) ,
\end{equation}

\noindent where
\begin{equation}
H_{1}\left( \varphi \right) =\int d\varphi \,\left( \sinh \left( \varphi
\right) ^{2}-a^{2}\right) \left[ 2\,\mu \,\sec h\left( \varphi \right)
+\left( \lambda -2\,\mu \right) \cosh \left( \varphi \right) \right] .
\end{equation}

Concluding the work, we should say that we are working on the extension of this
approach to the case of non-BPS states, the consequences for the appearing of bags,
junctions and networks of topological defects \cite {junction}, a more extens1ve
exploration of the new models coming from the approach proposed here, quantum
extensions of this classical method and to the case of a greater number of fields
\cite{clovis}

This work was partially supported by CNPq and FAPESP.

\end{document}